%% file: paper.tex
\documentclass[letterpaper,compsoc,twoside]{IEEEtran}
\usepackage{cmap} % fix search and cut-and-paste in Acrobat
\usepackage{ifthen}
\usepackage[T1]{fontenc}
\usepackage[utf8]{inputenc}
\usepackage{amsmath}
\usepackage{booktabs}
\usepackage{color}

\usepackage[font={small,it},labelfont=bf]{caption}
\usepackage{float}

\usepackage{longtable,ltcaption,array}
\newlength{\DUtablewidth} % internal use in tables

\usepackage{scipy}
\makeatletter
\def\PY@reset{\let\PY@it=\relax \let\PY@bf=\relax%
    \let\PY@ul=\relax \let\PY@tc=\relax%
    \let\PY@bc=\relax \let\PY@ff=\relax}
\def\PY@tok#1{\csname PY@tok@#1\endcsname}
\def\PY@toks#1+{\ifx\relax#1\empty\else%
    \PY@tok{#1}\expandafter\PY@toks\fi}
\def\PY@do#1{\PY@bc{\PY@tc{\PY@ul{%
    \PY@it{\PY@bf{\PY@ff{#1}}}}}}}
\def\PY#1#2{\PY@reset\PY@toks#1+\relax+\PY@do{#2}}

\@namedef{PY@tok@w}{\def\PY@tc##1{\textcolor[rgb]{0.73,0.73,0.73}{##1}}}
\@namedef{PY@tok@c}{\let\PY@it=\textit\def\PY@tc##1{\textcolor[rgb]{0.25,0.50,0.56}{##1}}}
\@namedef{PY@tok@cp}{\def\PY@tc##1{\textcolor[rgb]{0.00,0.44,0.13}{##1}}}
\@namedef{PY@tok@cs}{\def\PY@tc##1{\textcolor[rgb]{0.25,0.50,0.56}{##1}}\def\PY@bc##1{{\setlength{\fboxsep}{0pt}\colorbox[rgb]{1.00,0.94,0.94}{\strut ##1}}}}
\@namedef{PY@tok@k}{\let\PY@bf=\textbf\def\PY@tc##1{\textcolor[rgb]{0.00,0.44,0.13}{##1}}}
\@namedef{PY@tok@kp}{\def\PY@tc##1{\textcolor[rgb]{0.00,0.44,0.13}{##1}}}
\@namedef{PY@tok@kt}{\def\PY@tc##1{\textcolor[rgb]{0.56,0.13,0.00}{##1}}}
\@namedef{PY@tok@o}{\def\PY@tc##1{\textcolor[rgb]{0.40,0.40,0.40}{##1}}}
\@namedef{PY@tok@ow}{\let\PY@bf=\textbf\def\PY@tc##1{\textcolor[rgb]{0.00,0.44,0.13}{##1}}}
\@namedef{PY@tok@nb}{\def\PY@tc##1{\textcolor[rgb]{0.00,0.44,0.13}{##1}}}
\@namedef{PY@tok@nf}{\def\PY@tc##1{\textcolor[rgb]{0.02,0.16,0.49}{##1}}}
\@namedef{PY@tok@nc}{\let\PY@bf=\textbf\def\PY@tc##1{\textcolor[rgb]{0.05,0.52,0.71}{##1}}}
\@namedef{PY@tok@nn}{\let\PY@bf=\textbf\def\PY@tc##1{\textcolor[rgb]{0.05,0.52,0.71}{##1}}}
\@namedef{PY@tok@ne}{\def\PY@tc##1{\textcolor[rgb]{0.00,0.44,0.13}{##1}}}
\@namedef{PY@tok@nv}{\def\PY@tc##1{\textcolor[rgb]{0.73,0.38,0.84}{##1}}}
\@namedef{PY@tok@no}{\def\PY@tc##1{\textcolor[rgb]{0.38,0.68,0.84}{##1}}}
\@namedef{PY@tok@nl}{\let\PY@bf=\textbf\def\PY@tc##1{\textcolor[rgb]{0.00,0.13,0.44}{##1}}}
\@namedef{PY@tok@ni}{\let\PY@bf=\textbf\def\PY@tc##1{\textcolor[rgb]{0.84,0.33,0.22}{##1}}}
\@namedef{PY@tok@na}{\def\PY@tc##1{\textcolor[rgb]{0.25,0.44,0.63}{##1}}}
\@namedef{PY@tok@nt}{\let\PY@bf=\textbf\def\PY@tc##1{\textcolor[rgb]{0.02,0.16,0.45}{##1}}}
\@namedef{PY@tok@nd}{\let\PY@bf=\textbf\def\PY@tc##1{\textcolor[rgb]{0.33,0.33,0.33}{##1}}}
\@namedef{PY@tok@s}{\def\PY@tc##1{\textcolor[rgb]{0.25,0.44,0.63}{##1}}}
\@namedef{PY@tok@sd}{\let\PY@it=\textit\def\PY@tc##1{\textcolor[rgb]{0.25,0.44,0.63}{##1}}}
\@namedef{PY@tok@si}{\let\PY@it=\textit\def\PY@tc##1{\textcolor[rgb]{0.44,0.63,0.82}{##1}}}
\@namedef{PY@tok@se}{\let\PY@bf=\textbf\def\PY@tc##1{\textcolor[rgb]{0.25,0.44,0.63}{##1}}}
\@namedef{PY@tok@sr}{\def\PY@tc##1{\textcolor[rgb]{0.14,0.33,0.53}{##1}}}
\@namedef{PY@tok@ss}{\def\PY@tc##1{\textcolor[rgb]{0.32,0.47,0.09}{##1}}}
\@namedef{PY@tok@sx}{\def\PY@tc##1{\textcolor[rgb]{0.78,0.36,0.04}{##1}}}
\@namedef{PY@tok@m}{\def\PY@tc##1{\textcolor[rgb]{0.13,0.50,0.31}{##1}}}
\@namedef{PY@tok@gh}{\let\PY@bf=\textbf\def\PY@tc##1{\textcolor[rgb]{0.00,0.00,0.50}{##1}}}
\@namedef{PY@tok@gu}{\let\PY@bf=\textbf\def\PY@tc##1{\textcolor[rgb]{0.50,0.00,0.50}{##1}}}
\@namedef{PY@tok@gd}{\def\PY@tc##1{\textcolor[rgb]{0.63,0.00,0.00}{##1}}}
\@namedef{PY@tok@gi}{\def\PY@tc##1{\textcolor[rgb]{0.00,0.63,0.00}{##1}}}
\@namedef{PY@tok@gr}{\def\PY@tc##1{\textcolor[rgb]{1.00,0.00,0.00}{##1}}}
\@namedef{PY@tok@ge}{\let\PY@it=\textit}
\@namedef{PY@tok@gs}{\let\PY@bf=\textbf}
\@namedef{PY@tok@gp}{\let\PY@bf=\textbf\def\PY@tc##1{\textcolor[rgb]{0.78,0.36,0.04}{##1}}}
\@namedef{PY@tok@go}{\def\PY@tc##1{\textcolor[rgb]{0.20,0.20,0.20}{##1}}}
\@namedef{PY@tok@gt}{\def\PY@tc##1{\textcolor[rgb]{0.00,0.27,0.87}{##1}}}
\@namedef{PY@tok@err}{\def\PY@bc##1{{\setlength{\fboxsep}{\string -\fboxrule}\fcolorbox[rgb]{1.00,0.00,0.00}{1,1,1}{\strut ##1}}}}
\@namedef{PY@tok@kc}{\let\PY@bf=\textbf\def\PY@tc##1{\textcolor[rgb]{0.00,0.44,0.13}{##1}}}
\@namedef{PY@tok@kd}{\let\PY@bf=\textbf\def\PY@tc##1{\textcolor[rgb]{0.00,0.44,0.13}{##1}}}
\@namedef{PY@tok@kn}{\let\PY@bf=\textbf\def\PY@tc##1{\textcolor[rgb]{0.00,0.44,0.13}{##1}}}
\@namedef{PY@tok@kr}{\let\PY@bf=\textbf\def\PY@tc##1{\textcolor[rgb]{0.00,0.44,0.13}{##1}}}
\@namedef{PY@tok@bp}{\def\PY@tc##1{\textcolor[rgb]{0.00,0.44,0.13}{##1}}}
\@namedef{PY@tok@fm}{\def\PY@tc##1{\textcolor[rgb]{0.02,0.16,0.49}{##1}}}
\@namedef{PY@tok@vc}{\def\PY@tc##1{\textcolor[rgb]{0.73,0.38,0.84}{##1}}}
\@namedef{PY@tok@vg}{\def\PY@tc##1{\textcolor[rgb]{0.73,0.38,0.84}{##1}}}
\@namedef{PY@tok@vi}{\def\PY@tc##1{\textcolor[rgb]{0.73,0.38,0.84}{##1}}}
\@namedef{PY@tok@vm}{\def\PY@tc##1{\textcolor[rgb]{0.73,0.38,0.84}{##1}}}
\@namedef{PY@tok@sa}{\def\PY@tc##1{\textcolor[rgb]{0.25,0.44,0.63}{##1}}}
\@namedef{PY@tok@sb}{\def\PY@tc##1{\textcolor[rgb]{0.25,0.44,0.63}{##1}}}
\@namedef{PY@tok@sc}{\def\PY@tc##1{\textcolor[rgb]{0.25,0.44,0.63}{##1}}}
\@namedef{PY@tok@dl}{\def\PY@tc##1{\textcolor[rgb]{0.25,0.44,0.63}{##1}}}
\@namedef{PY@tok@s2}{\def\PY@tc##1{\textcolor[rgb]{0.25,0.44,0.63}{##1}}}
\@namedef{PY@tok@sh}{\def\PY@tc##1{\textcolor[rgb]{0.25,0.44,0.63}{##1}}}
\@namedef{PY@tok@s1}{\def\PY@tc##1{\textcolor[rgb]{0.25,0.44,0.63}{##1}}}
\@namedef{PY@tok@mb}{\def\PY@tc##1{\textcolor[rgb]{0.13,0.50,0.31}{##1}}}
\@namedef{PY@tok@mf}{\def\PY@tc##1{\textcolor[rgb]{0.13,0.50,0.31}{##1}}}
\@namedef{PY@tok@mh}{\def\PY@tc##1{\textcolor[rgb]{0.13,0.50,0.31}{##1}}}
\@namedef{PY@tok@mi}{\def\PY@tc##1{\textcolor[rgb]{0.13,0.50,0.31}{##1}}}
\@namedef{PY@tok@il}{\def\PY@tc##1{\textcolor[rgb]{0.13,0.50,0.31}{##1}}}
\@namedef{PY@tok@mo}{\def\PY@tc##1{\textcolor[rgb]{0.13,0.50,0.31}{##1}}}
\@namedef{PY@tok@ch}{\let\PY@it=\textit\def\PY@tc##1{\textcolor[rgb]{0.25,0.50,0.56}{##1}}}
\@namedef{PY@tok@cm}{\let\PY@it=\textit\def\PY@tc##1{\textcolor[rgb]{0.25,0.50,0.56}{##1}}}
\@namedef{PY@tok@cpf}{\let\PY@it=\textit\def\PY@tc##1{\textcolor[rgb]{0.25,0.50,0.56}{##1}}}
\@namedef{PY@tok@c1}{\let\PY@it=\textit\def\PY@tc##1{\textcolor[rgb]{0.25,0.50,0.56}{##1}}}

\makeatother

\providecommand*{\DUrole}[2]{%
  \ifcsname docutilsrole#1\endcsname%
    \csname docutilsrole#1\endcsname{#2}%
  \else
    \csname DUrole#1\endcsname{#2}%
  \fi%
}

\ifthenelse{\isundefined{\hypersetup}}{
  \usepackage[colorlinks=true,linkcolor=blue,urlcolor=blue]{hyperref}
  \usepackage{bookmark}
  \urlstyle{same} % normal text font (alternatives: tt, rm, sf)
}{}

\begin{document}
\newcounter{footnotecounter}\title{Experience report of physics-informed neural networks in fluid simulations: pitfalls and frustration}\author{Pi-Yueh Chuang$^{\setcounter{footnotecounter}{3}\fnsymbol{footnotecounter}\setcounter{footnotecounter}{1}\fnsymbol{footnotecounter}}$%
          \setcounter{footnotecounter}{1}\thanks{\fnsymbol{footnotecounter} %
          Corresponding author: \protect\href{mailto:pychuang@gwu.edu}{pychuang@gwu.edu}}\setcounter{footnotecounter}{3}\thanks{\fnsymbol{footnotecounter} Department of Mechanical and Aerospace Engineering, The George Washington University, Washington, DC 20052, USA}, Lorena A. Barba$^{\setcounter{footnotecounter}{3}\fnsymbol{footnotecounter}}$\thanks{%

          \noindent%
          Copyright\,\copyright\,2022 Pi-Yueh Chuang et al. This is an open-access article distributed under the terms of the Creative Commons Attribution License, which permits unrestricted use, distribution, and reproduction in any medium, provided the original author and source are credited.%
        }}\maketitle
          \renewcommand{\leftmark}{PROC. OF THE 21st PYTHON IN SCIENCE CONF. (SCIPY 2022)}
          \renewcommand{\rightmark}{EXPERIENCE REPORT OF PHYSICS-INFORMED NEURAL NETWORKS IN FLUID SIMULATIONS: PITFALLS AND FRUSTRATION}
        
\InputIfFileExists{page_numbers.tex}{}{}
\newcommand*{\docutilsroleref}{\ref}
\newcommand*{\docutilsrolelabel}{\label}
\newcommand*\DUrolecode[1]{#1}
\providecommand*\DUrolecite[1]{\cite{#1}}
\begin{abstract}Though PINNs (physics-informed neural networks) are now deemed as a complement to traditional CFD (computational fluid dynamics) solvers rather than a replacement, their ability to solve the Navier-Stokes equations without given data is still of great interest.
This report presents our not-so-successful experiments of solving the Navier-Stokes equations with PINN as a replacement to traditional solvers.
We aim to, with our experiments, prepare readers for the challenges they may face if they are interested in data-free PINN.
In this work, we used two standard flow problems: 2D Taylor-Green vortex at $Re = 100$ and 2D cylinder flow at $Re = 200$.
The PINN method solved the 2D Taylor-Green vortex problem with acceptable results, and we used this flow as an accuracy and performance benchmark.
About 32 hours of training were required for the PINN method's accuracy to match the accuracy of a $16 \times 16$ finite-difference simulation, which took less than 20 seconds.
The 2D cylinder flow, on the other hand, did not produce a physical solution.
The PINN method behaved like a steady-flow solver and did not capture the vortex shedding phenomenon.
By sharing our experience, we would like to emphasize that the PINN method is still a work-in-progress, especially in terms of solving flow problems without any given data.
More work is needed to make PINN feasible for real-world problems in such applications.
(Reproducibility package: \DUrole{cite}{pi_yueh_chuang_2022_6592457}.)\end{abstract}

\subsection{1. Introduction%
  \label{introduction}%
}

Recent advances in computing and programming techniques have motivated practitioners to revisit deep learning applications in computational fluid dynamics (CFD).
We use the verb \textquotedbl{}revisit\textquotedbl{} because deep learning applications in CFD already existed going back to at least the 1990s,
for example, using neural networks as surrogate models (\DUrole{cite}{Linse1993,Faller1997}).
Another example is the work of Lagaris and his/her colleagues (\DUrole{cite}{lagaris_artificial_1998}) on solving partial differential equations with fully-connected neural networks back in 1998.
Similar work with radial basis function networks can be found in reference \DUrole{cite}{Li2003}.
Nevertheless, deep learning applications in CFD did not get much attention until this decade, thanks to modern computing technology, including GPUs, cloud computing, high-level libraries like PyTorch and TensorFlow, and their Python APIs.

Solving partial differential equations with deep learning is particularly interesting to CFD researchers and practitioners.
The PINN (physics-informed neural network) method denotes an approach to incorporate deep learning in CFD applications, where solving partial differential equations plays the key role.
These partial differential equations include the well-known Navier-Stokes equations—one of the Millennium Prize Problems.
The universal approximation theorem (\DUrole{cite}{hornik_approximation_1991}) implies that neural networks can model the solution to the Navier-Stokes equations with high fidelity and capture complicated flow details as long as networks are big enough.
The idea of PINN methods can be traced back to \DUrole{cite}{dissanayake_neural-network-based_1994}, while the name PINN was coined in \DUrole{cite}{raissi_physics-informed_2019}.
Human-provided data are not necessary in applying PINN \DUrole{cite}{lu_deepxde:_2021}, making it a potential alternative to traditional CFD solvers.
Sometimes it is branded as unsupervised learning—it does not rely on human-provided data, making it sound very \emph{\textquotedbl{}AI.\textquotedbl{}}
It is now common to see headlines like \emph{\textquotedbl{}AI has cracked the Navier-Stokes equations\textquotedbl{}} in recent popular science articles (\DUrole{cite}{hao_ai_2020}).

Though data-free PINN as an alternative to traditional CFD solvers may sound attractive,
PINN can also be used under data-driven configurations, for which it is better suited.
Cai et al. \DUrole{cite}{cai_physics-informed_2021} state that PINN is not meant to be a replacement of existing CFD solvers due to its inferior accuracy and efficiency.
The most useful applications of PINN should be those with some given data, and thus the models are trained against the data.
For example, when we have experimental measurements or partial simulation results (coarse-grid data, limited numbers of snapshots, etc.) from traditional CFD solvers, PINN may be useful to reconstruct the flow or to be a surrogate model.

Nevertheless, data-free PINN may offer some advantages over traditional solvers, and using data-free PINN to replace traditional solvers is still of great interest to researchers (e.g., \DUrole{cite}{karali_novel_2021}).
First, it is a mesh-free scheme, which benefits engineering problems where fluid flows interact with objects of complicated geometries.
Simulating these fluid flows with traditional numerical methods usually requires high-quality unstructured meshes
with time-consuming human intervention in the pre-processing stage before actual simulations.
The second benefit of PINN is that the trained models approximate the governing equations' general solutions, meaning there is no need to solve the equations repeatedly for different flow parameters.
For example, a flow model taking boundary velocity profiles as its input arguments can predict flows under different boundary velocity profiles after training.
Conventional numerical methods, on the contrary, require repeated simulations, each one covering one boundary velocity profile.
This feature could help in situations like engineering design optimization:
the process of running sets of experiments to conduct parameter sweeps and find the optimal values or geometries for products.
Given these benefits, researchers continue studying and improving the usability of data-free PINN (e.g., \DUrole{cite}{wang_when_2022,du_evolutional_2021,wang_understanding_2021,sirignano_dgm:_2018}).

Data-free PINN, however, is not ready nor meant to replace traditional CFD solvers.
This claim may be obvious to researchers experienced in PINN, but it may not be clear to others, especially to CFD end-users without ample expertise in numerical methods.
Even in literature that aims to improve PINN, it's common to see only the success stories with simple CFD problems.
Important information concerning the feasibility of PINN in practical and real-world applications is often missing from these success stories.
For example, few reports discuss the required computing resources, the computational cost of training, the convergence properties, or the error analysis of PINN.
PINN suffers from performance and solvability issues due to the need for high-order automatic differentiation and multi-objective nonlinear optimization.
Evaluating high-order derivatives using automatic differentiation increases the computational graphs of neural networks.
And multi-objective optimization, which reduces all the residuals of the differential equations, initial conditions, and boundary conditions, makes the training difficult to converge to small-enough loss values.
Fluid flows are sensitive nonlinear dynamical systems in which a small change or error in inputs may produce a very different flow field.
So to get correct solutions, the optimization in PINN needs to minimize the loss to values very close to zero, further compromising the method's solvability and performance.

This paper reports on our not-so-successful PINN story as a lesson learned to readers, so they can be aware of the challenges they may face if they consider using data-free PINN in real-world applications.
Our story includes two computational experiments as case studies to benchmark the PINN method's accuracy and computational performance.
The first case study is a Taylor-Green vortex, solved successfully though not to our complete satisfaction.
We will discuss the performance of PINN using this case study.
The second case study, flow over a cylinder, did not even result in a physical solution.
We will discuss the frustration we encountered with PINN in this case study.

We built our PINN solver with the help of NVIDIA's Modulus library (\DUrole{cite}{noauthor_modulus_nodate}).
Modulus is a high-level Python package built on top of PyTorch that helps users develop PINN-based differential equation solvers.
Also, in each case study, we also carried out simulations with our CFD solver, PetIBM (\DUrole{cite}{chuang_petibm_2018}).
PetIBM is a traditional solver using staggered-grid finite difference methods with MPI parallelization and GPU computing.
PetIBM simulations in each case study served as baseline data.
For all cases, configurations, post-processing scripts, and required Singularity image definitions can be found at reference \DUrole{cite}{pi_yueh_chuang_2022_6592457}.

This paper is structured as follows: the second section briefly describes the PINN method and an analogy to traditional CFD methods.
The third and fourth sections provide our computational experiments of the Taylor-Green vortex in 2D and a 2D laminar cylinder flow with vortex shedding.
Most discussions happen in the corresponding case studies.
The last section presents the conclusion and discussions that did not fit into either one of the cases.

\subsection{2. Solving Navier-Stokes equations with PINN%
  \label{solving-navier-stokes-equations-with-pinn}%
}

The incompressible Navier-Stokes equations in vector form are composed of the continuity equation:\begin{equation}
\label{eq:continuity}
\nabla \cdot\vec{U}=0
\end{equation}and momentum equations:\begin{equation}
\label{eq:momentum}
\frac{\partial \vec{U}}{\partial t}+(\vec{U} \cdot \nabla) \vec{U}=-\frac{1}{\rho} \nabla p +\nu \nabla^{2} \vec{U} + \vec{g}
\end{equation}where $\rho=\rho(\vec{x}, t)$, $\nu=\nu(\vec{x}, t)$, and $p=p(\vec{x}, t)$ are scalar fields denoting density, kinematic viscosity, and pressure, respectively.
$\vec{x}$ denotes the spatial coordinate, and $\vec{x}=\left[x,\ y\right]^{\mathsf{T}}$ in two dimensions.
The density and viscosity fields are usually known and given, while the pressure field is unknown.
$\vec{U}=\vec{U}(\vec{x}, t)=\left[u(x, y, t),\ v(x, y, t)\right]^\mathsf{T}$ is a vector field for flow velocity.
All of them are functions of the spatial coordinate in the computational domain $\Omega$ and time before a given limit $T$.
The gravitational field $\vec{g}$ may also be a function of space and time, though it is usually a constant.
A solution to the Navier-Stokes equations is subjected to an initial condition and boundary conditions:\begin{equation}
\label{eq:ic-and-bc}
\left\{
   \begin{array}{lll}
      \vec{U}(\vec{x}, t)=\vec{U}_0(\vec{x}), & \forall \vec{x} \in \Omega, & t=0 \\
      \vec{U}(\vec{x}, t)=\vec{U}_\Gamma(\vec{x}, t), & \forall \vec{x} \in \Gamma, & t\in [0, T] \\
      p(\vec{x}, t)=p_\Gamma(x, t), & \forall \vec{x} \in \Gamma, & t \in [0, T]
   \end{array}
\right.
\end{equation}where $\Gamma$ represents the boundary of the computational domain.

\subsubsection{2.1. The PINN method%
  \label{the-pinn-method}%
}

The basic form of the PINN method (\DUrole{cite}{raissi_physics-informed_2019,cai_physics-informed_2021}) starts from approximating $\vec{U}$ and $p$ with a neural network:\begin{equation}
\label{eq:neural-network}
\begin{bmatrix}
\vec{U} \\ p
\end{bmatrix}(\vec{x}, t)
\approx
G(\vec{x}, t; \Theta)
\end{equation}Here we use a single network that predicts both pressure and velocity fields.
It is also possible to use different networks for them separately.
Later in this work, we will use $G^U$ and $G^p$ to denote the predicted velocity and pressure from the neural network.
$\Theta$ at this point represents the free parameters of the network.

To determine the free parameters, $\Theta$, ideally, we hope the approximate solution gives zero residuals for equations (\DUrole{ref}{eq:continuity}), (\DUrole{ref}{eq:momentum}), and (\DUrole{ref}{eq:ic-and-bc}).
That is\begin{equation}
\label{eq:residuals}
\begin{aligned}
   & r_{1}(\vec{x}, t; \Theta) \equiv \nabla \cdot G^{U} = 0 \\
   & r_{2}(\vec{x}, t; \Theta) \equiv \frac{\partial G^{U}}{\partial t}+(G^{U} \cdot \nabla) G^{U}+\frac{1}{\rho} \nabla G^p -\nu \nabla^{2} G^{U} - \vec{g} =0 \\
   & r_{3}(\vec{x}; \Theta) \equiv G^{U}_{t=0}-\vec{U}_0 = 0 \\
   & r_{4}(\vec{x}, t; \Theta) \equiv G^{U}-\vec{U}_\Gamma = 0,\ \forall \vec{x} \in \Gamma \\
   & r_{5}(\vec{x}, t; \Theta) \equiv G^{p}-p_\Gamma = 0,\ \forall \vec{x} \in \Gamma \\
\end{aligned}
\end{equation}And the set of desired parameter, $\Theta=\theta$, is the common zero root of all the residuals.

The derivatives of $G$ with respect to $\vec{x}$ and $t$ are usually obtained using automatic differentiation.
Nevertheless, it is possible to use analytical derivatives when the chosen network architecture is simple enough, as reported by early-day literature (\DUrole{cite}{lagaris_artificial_1998,Li2003}).

If residuals in (\DUrole{ref}{eq:residuals}) are not complicated, and if the number of the parameters, $N_\Theta$, is small enough, we may numerically find the zero root by solving a system of $N_\Theta$ nonlinear equations generated from a suitable set of $N_\Theta$ spatial-temporal points.
However, the scenario rarely happens as $G$ is usually highly complicated and $N_\Theta$ is large.
Moreover, we do not even know if such a zero root exists for the equations in (\DUrole{ref}{eq:residuals}).

Instead, in PINN, the condition is relaxed.
We do not seek the zero root of (\DUrole{ref}{eq:residuals}) but just hope to find a set of parameters that make the residuals sufficiently close to zero.
Consider the sum of the $l_2$ norms of residuals:\begin{equation}
\label{eq:total-residual}
r(\vec{x}, t; \Theta=\theta) \equiv \sum\limits_{i=1}^{5} \lVert r_i(\vec{x}, t; \Theta=\theta) \rVert^2,\ \forall \left\{\begin{array}{l}x \in \Omega \\ t\in[0, T]\end{array}\right.
\end{equation}The $\theta$ that makes residuals closest to zero (or even equal to zero if such $\theta$ exists) also makes (\DUrole{ref}{eq:total-residual}) minimal because $r(\vec{x}, t; \Theta) \ge 0$.
In other words,\begin{equation}
\label{eq:objective}
\theta = \operatorname*{arg\,min}\limits_{\Theta} r(\vec{x}, t; \Theta)\,\ \forall \left\{\begin{array}{l}x \in \Omega \\ t\in[0, T]\end{array}\right.
\end{equation}This poses a fundamental difference between the PINN method and traditional CFD schemes, making it potentially more difficult for the PINN method to achieve the same accuracy as the traditional schemes.
We will discuss this more in section 3.
Note that in practice, each loss term on the right-hand-side of equation (\DUrole{ref}{eq:total-residual}) is weighted.
We ignore the weights here for demonstrating purpose.

To solve (\DUrole{ref}{eq:objective}), theoretically, we can use any number of spatial-temporal points, which eases the need of computational resources, compared to finding the zero root directly.
Gradient-descent-based optimizers further reduce the computational cost, especially in terms of memory usage and the difficulty of parallelization.
Alternatively, Quasi-Newton methods may work but only when $N_\Theta$ is small enough.

However, even though equation (\DUrole{ref}{eq:objective}) may be solvable, it is still a significantly expensive task.
While typical data-driven learning requires one back-propagation pass on the derivatives of the loss function, here automatic differentiation is needed to evaluate the derivatives of $G$ with respect to $\vec{x}$ and $t$.
The first-order derivatives require one back-propagation on the network, while the second-order derivatives present in the diffusion term $\nabla^2 G^U$ require an additional back-propagation on the first-order derivatives' computational graph.
Finally, to update parameters in an optimizer, the gradients of $G$ with respect to parameters $\Theta$ requires another back-propagation on the graph of the second-order derivatives.
This all leads to a very large computational graph.
We will see the performance of the PINN method in the case studies.

In summary, when viewing the PINN method as supervised machine learning, the inputs of a network are spatial-temporal coordinates, and the outputs are the physical quantities of our interest.
The loss or objective functions in PINN are governing equations that regulate how the target physical quantities should behave.
The use of governing equations eliminates the need for true answers.
A trivial example is using Bernoulli's equation as the loss function, i.e., $loss=\frac{u^2}{2g}+\frac{p}{\rho g}-H_0+z(x)$, and a neural network predicts the flow speed $u$ and pressure $p$ at a given location $x$ along a streamline.
(The gravitational acceleration $g$, density $\rho$, energy head $H_0$, and elevation $z(x)$ are usually known and given.)
Such a loss function regulates the relationship between predicted $u$ and $p$ and does not need true answers for the two quantities.
Unlike Bernoulli’s equation, most governing equations in physics are usually differential equations (e.g., heat equations).
The main difference is that now the PINN method needs automatic differentiation to evaluate the loss.
Regardless of the forms of governing equations, spatial-temporal coordinates are the only data required during training.
Hence, throughout this paper, training data means spatial-temporal points and does not involve any true answers to predicted quantities.
(Note in some literature, the PINN method is applied to applications that do need true answers, see \DUrole{cite}{cai_physics-informed_2021}. These applications are out of scope here.)

\subsubsection{2.2. An analogy to conventional numerical methods%
  \label{an-analogy-to-conventional-numerical-methods}%
}

For readers with a background in numerical methods for partial differential equations, we would like to make an analogy between traditional numerical methods and PINN.

In obtaining strong solutions to differential equations, we can describe the solution workflows of most numerical methods with five stages:
\begin{enumerate}
\item 

\emph{Designing the approximate solution with undetermined parameters}
\item 

\emph{Choosing proper approximation for derivatives}
\item 

\emph{Obtaining the so-called modified equation by substituting approximate derivatives into the differential equations and initial/boundary conditions}
\item 

\emph{Generating a system of linear/nonlinear algebraic equations}
\item 

\emph{Solving the system of equations}\end{enumerate}

For example, to solve $\nabla U^2(x)=s(x)$, the most naive spectral method (\DUrole{cite}{trefethen_spectral_2000}) approximates the solution with $U(x)\approx G(x)=\sum\limits_{i=1}^{N}c_i\phi_i(x)$, where $c_i$ represents undetermined parameters, and $\phi_i(x)$ denotes a set of either polynomials, trigonometric functions, or complex exponentials.
Next, obtaining the first derivative of $U$ is straightforward—we can just assume $U^{\prime}(x)\approx G^{\prime}(x)=\sum\limits_{i=1}^{N}c_i \phi_i^{\prime}(x)$.
The second-order derivative may be more tricky.
One can assume $U^{\prime\prime}(x)\approx G^{\prime\prime}=\sum\limits_{i=1}^{N}c_i \phi_i^{\prime\prime}(x)$.
Or, another choice for nodal bases (i.e., when $\phi_i(x)$ is chosen to make $c_i\equiv G(x_i)$) is $U^{\prime\prime}(x)\approx \sum\limits_{i=1}^{N}c_i G^{\prime}(x_i)$.
Because $\phi_i(x)$ is known, the derivatives are analytical.
After substituting the approximate solution and derivatives in to the target differential equation, we need to solve for parameters $c_1,\cdots,c_N$.
We do so by selecting $N$ points from the computational domain and creating a system of $N$ linear equations:\begin{equation}
\label{eq:spectral-linear-sys}
\begin{bmatrix}
\phi_1^{\prime\prime}(x_1) & \cdots & \phi_N^{\prime\prime}(x_1) \\
\vdots & \ddots & \vdots \\
\phi_1^{\prime\prime}(x_N) & \cdots & \phi_N^{\prime\prime}(x_N)
\end{bmatrix}
\begin{bmatrix}
c_1 \\ \vdots \\ c_N
\end{bmatrix}
-
\begin{bmatrix}
s(x_1) \\ \vdots \\ s(x_N)
\end{bmatrix}
= 0
\end{equation}Finally, we determine the parameters by solving this linear system.
Though this example uses a spectral method, the workflow also applies to many other numerical methods, such as finite difference methods, which can be reformatted as a form of spectral method.

With this workflow in mind, it should be easy to see the analogy between PINN and conventional numerical methods.
Aside from using much more complicated approximate solutions, the major difference lies in how to determine the unknown parameters in the approximate solutions.
While traditional methods solve the zero-residual conditions, PINN relies on searching the minimal residuals.
A secondary difference is how to approximate derivatives.
Conventional numerical methods use analytical or numerical differentiation of the approximate solutions, and the PINN methods usually depends on automatic differentiation.
This difference may be minor as we are still able to use analytical differentiation for simple network architectures with PINN.
However, automatic differentiation is a major factor affecting PINN's performance.

\subsection{3. Case 1: Taylor-Green vortex: accuracy and performance%
  \label{case-1-taylor-green-vortex-accuracy-and-performance}%
}

\subsubsection{3.1. 2D Taylor-Green vortex%
  \label{d-taylor-green-vortex}%
}

The Taylor-Green vortex represents a family of flows with a specific form of analytical initial flow conditions in both 2D and 3D.
The 2D Taylor-Green vortex has closed-form analytical solutions with periodic boundary conditions, and hence they are standard benchmark cases for verifying CFD solvers.
In this work, we used the following 2D Taylor-Green vortex:\begin{equation}
\label{eq:tgv}
\left\{
\begin{aligned}
u(x, y, t) &= V_0\cos(\frac{x}{L})\sin(\frac{y}{L})\exp(-2\frac{\nu}{L^2}t) \\
v(x, y, t) &= - V_0 \sin(\frac{x}{L})\cos(\frac{y}{L})\exp(-2\frac{\nu}{L^2}t) \\
p(x, y, t) &= -\frac{\rho}{4}V_0^2\left(cos(\frac{2x}{L}) + cos(\frac{2y}{L})\right)\exp(-4\frac{\nu}{L^2}t) \\
\end{aligned}
\right.
\end{equation}where $V_0$ represents the peak (and also the lowest) velocity at $t=0$.
Other symbols carry the same meaning as those in section 2.

The periodic boundary conditions were applied to $x=-L\pi$, $x=L\pi$, $y=-L\pi$, and $y=L\pi$.
We used the following parameters in this work: $V_0=L=\rho=1.0$ and $\nu=0.01$.
These parameters correspond to Reynolds number $Re=100$. Figure \DUrole{ref}{fig:tgv-petibm-contour-t32} shows a snapshot of velocity at $t=32$.\begin{figure}[]\noindent\makebox[\columnwidth][c]{\includegraphics[width=\columnwidth]{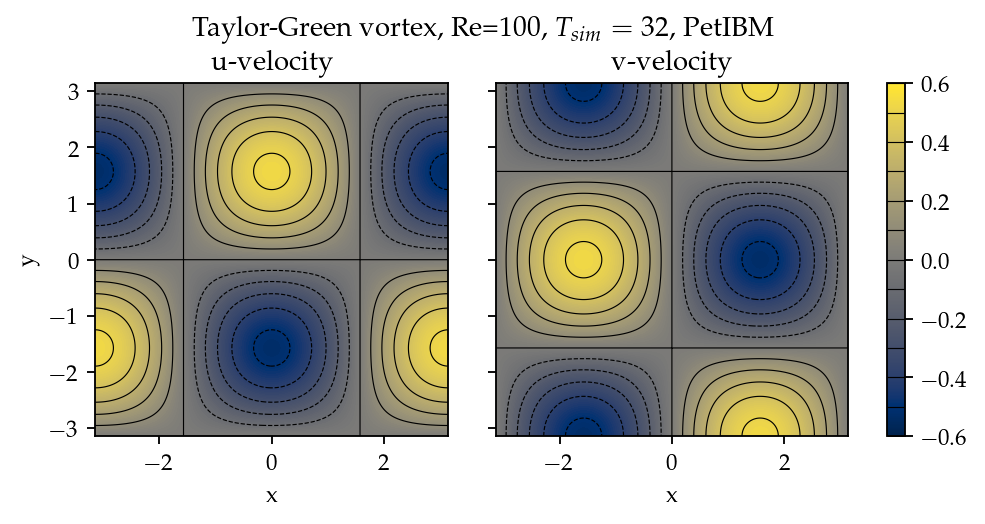}}
\caption{Contours of $u$ and $v$ at $t=32$ to demonstrate the solution of 2D Taylor-Green vortex. \DUrole{label}{fig:tgv-petibm-contour-t32}}
\end{figure}

\subsubsection{3.2. Solver and runtime configurations%
  \label{solver-and-runtime-configurations}%
}

The neural network used in the PINN solver is a fully-connected neural network with 6 hidden layers and 256 neurons per layer.
The activation functions are SiLU (\DUrole{cite}{hendrycks_gaussian_2016}).
We used Adam for optimization, and its initial parameters are the defaults from PyTorch.
The learning rate exponentially decayed through PyTorch's \texttt{\DUrole{code}{ExponentialLR}} with \texttt{\DUrole{code}{gamma}} equal to $0.95^{1/10000}$.
Note we did not conduct hyperparameter optimization, given the computational cost.
The hyperparameters are mostly the defaults used by the 3D Taylor-Green example in Modulus (\DUrole{cite}{noauthor_modulus_nodate}).

The training data were simply spatial-temporal coordinates.
Before the training, the PINN solver pre-generated 18,432,000 spatial-temporal points to evaluate the residuals of the Navier-Stokes equations (the $r_1$ and $r_2$ in equation (\DUrole{ref}{eq:residuals})).
These training points were randomly chosen from the spatial domain $[-\pi, \pi]\times[-\pi, \pi]$ and temporal domain $(0, 100]$.
The solver used only 18,432 points in each training iteration, making it a batch training.
For the residual of the initial condition (the $r_3$), the solver also pre-generated 18,432,000 random spatial points and used only 18,432 per iteration.
Note that for $r_3$, the points were distributed in space only because $t=0$ is a fixed condition.
Because of the periodic boundary conditions, the solver did not require any training points for $r_4$ and $r_5$.

The hardware used for the PINN solver was a single node of NVIDIA's DGX-A100.
It was equipped with 8 A100 GPUs (80GB variants).
We carried out the training using different numbers of GPUs to investigate the performance of the PINN solver.
All cases were trained up to 1 million iterations.
Note that the parallelization was done with weak scaling, meaning increasing the number of GPUs would not reduce the workload of each GPU.
Instead, increasing the number of GPUs would increase the total and per-iteration numbers of training points.
Therefore, our expected outcome was that all cases required about the same wall time to finish, while the residual from using 8 GPUs would converge the fastest.

After training, the PINN solver's prediction errors (i.e., accuracy) were evaluated on cell centers of a $512 \times 512$ Cartesian mesh against the analytical solution.
With these spatially distributed errors, we calculated the $L_2$ error norm for a given $t$:\begin{equation}
\label{eq:l2norm}
L_2 = \sqrt{\int\limits_{\Omega} error(x, y)^2 \mathrm{d}\Omega} \approx \sqrt{\sum\limits_{i}\sum\limits_{j} error_{i, j}^2 \Delta \Omega_{i, j}}
\end{equation}where $i$ and $j$ here are the indices of a cell center in the Cartesian mesh. $\Delta\Omega_{i,j}$ is the corresponding cell area, $4\pi^2/512^2$ in this case.

We compared accuracy and performance against results using PetIBM.
All PetIBM simulations in this section were done with 1 K40 GPU and 6 CPU cores (Intel i7-5930K) on our old lab workstation.
We carried out 7 PetIBM simulations with different spatial resolutions: $2^k\times 2^k$ for $k=4, 5, \dots, 10$.
The time step size for each spatial resolution was $\Delta t=0.1/2^{k-4}$.

A special note should be made here: the PINN solver used single-precision floats, while PetIBM used double-precision floats.
It might sound unfair.
However, this discrepancy does not change the qualitative findings and conclusions, as we will see later.

\subsubsection{3.3. Results%
  \label{results}%
}

Figure \DUrole{ref}{fig:tgv-pinn-training-convergence} shows the convergence history of the total residuals (equation (\DUrole{ref}{eq:total-residual})).
Using more GPUs in weak scaling (i.e., more training points) did not accelerate the convergence, contrary to what we expected.
All cases converged at a similar rate.
Though without a quantitative criterion or justification, we considered that further training would not improve the accuracy.
Figure \DUrole{ref}{fig:tgv-pinn-contour-t32} gives a visual taste of what the predictions from the neural network look like.\begin{figure}[]\noindent\makebox[\columnwidth][c]{\includegraphics[width=\columnwidth]{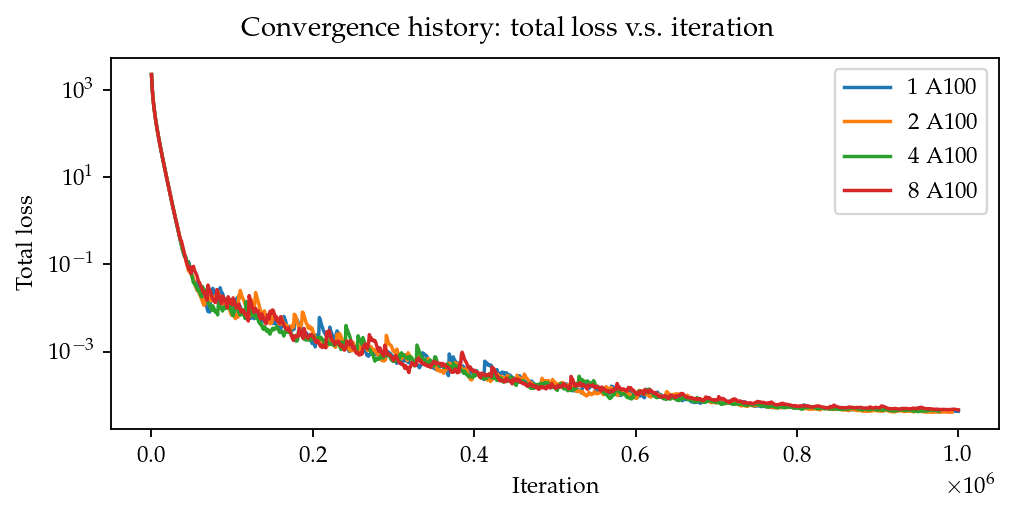}}
\caption{Total residuals (loss) with respect to training iterations. \DUrole{label}{fig:tgv-pinn-training-convergence}}
\end{figure}\begin{figure}[]\noindent\makebox[\columnwidth][c]{\includegraphics[width=\columnwidth]{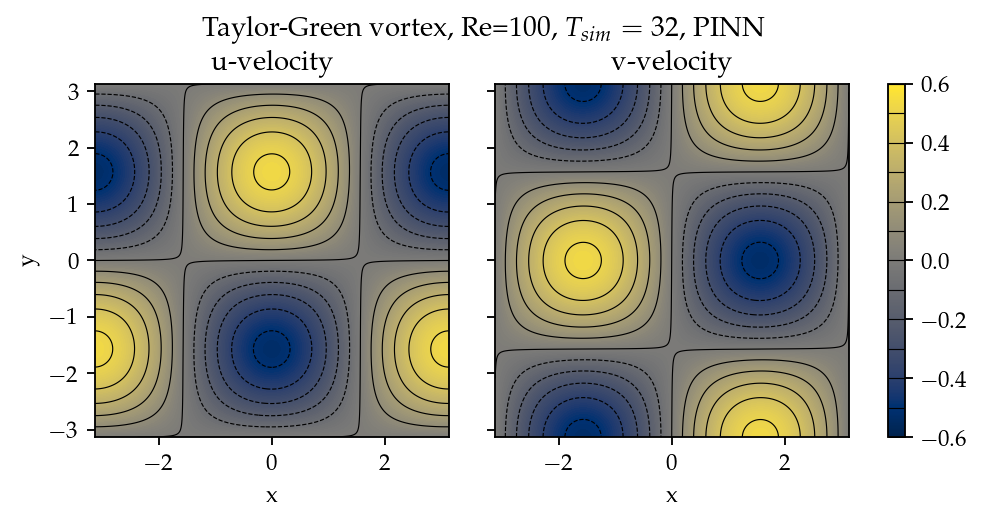}}
\caption{Contours of $u$ and $v$ at $t=32$ from the PINN solver. \DUrole{label}{fig:tgv-pinn-contour-t32}}
\end{figure}

The result visually agrees with that in figure \DUrole{ref}{fig:tgv-petibm-contour-t32}.
However, as shown in figure \DUrole{ref}{fig:tgv-sim-time-errors}, the error magnitudes from the PINN solver are much higher than those from PetIBM.
Figure \DUrole{ref}{fig:tgv-sim-time-errors} shows the prediction errors with respect to $t$.
We only present the error on the $u$ velocity as those for $v$ and $p$ are similar.
The accuracy of the PINN solver is similar to that of the $16 \times 16$ simulation with PetIBM.
Using more GPUs, which implies more training points, does not improve the accuracy.

Regardless of the magnitudes, the trends of the errors with respect to $t$ are similar for both PINN and PetIBM.
For PetIBM, the trend shown in figure \DUrole{ref}{fig:tgv-sim-time-errors} indicates that the temporal error is bounded, and the scheme is stable.
However, this concept does not apply to PINN as it does not use any time-marching schemes.
What this means for PINN is still unclear to us.
Nevertheless, it shows that PINN is able to propagate the influence of initial conditions to later times, which is a crucial factor for solving hyperbolic partial differential equations.

Figure \DUrole{ref}{fig:tgv-run-time-errors} shows the computational cost of PINN and PetIBM in terms of the desired accuracy versus the required wall time.
We only show the PINN results of 8 A100 GPUs on this figure.
We believe this type of plot may help evaluate the computational cost in engineering applications.
According to the figure, for example, achieving an accuracy of $10^{-3}$ at $t=2$ requires less than 1 second for PetIBM with 1 K40 and 6 CPU cores, but it requires more than 8 hours for PINN with at least 1 A100 GPU.\begin{figure}[]\noindent\makebox[\columnwidth][c]{\includegraphics[width=\columnwidth]{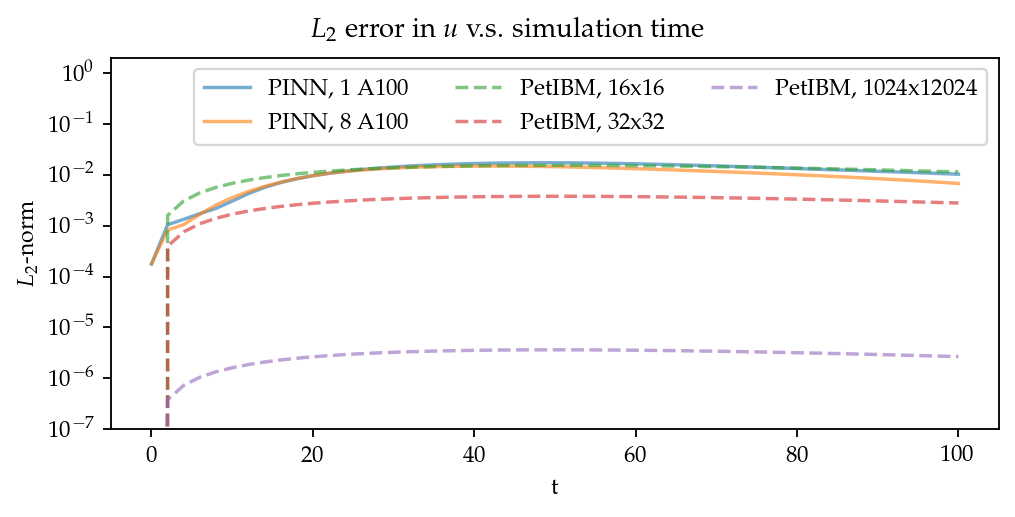}}
\caption{$L_2$ error norm versus simulation time. \DUrole{label}{fig:tgv-sim-time-errors}}
\end{figure}\begin{figure}[]\noindent\makebox[\columnwidth][c]{\includegraphics[width=\columnwidth]{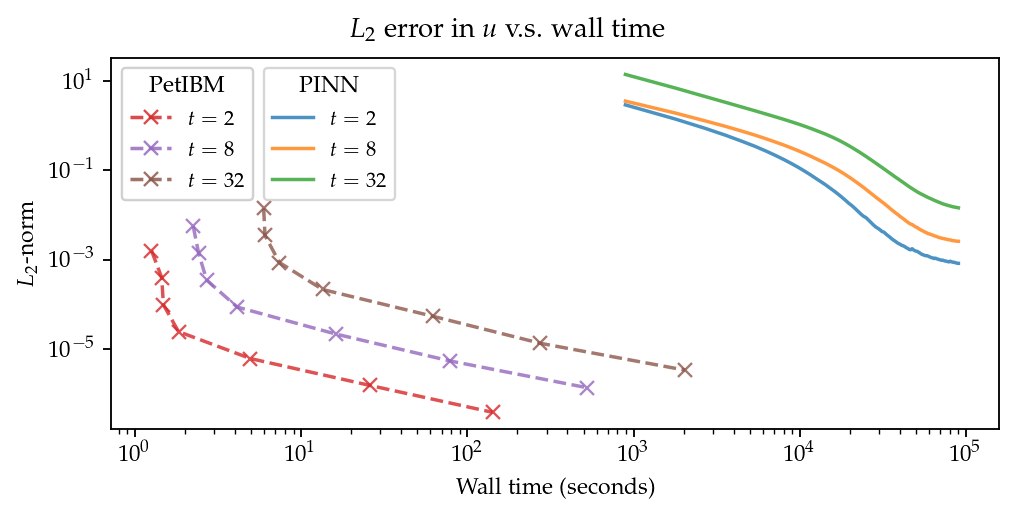}}
\caption{$L_2$ error norm versus wall time. \DUrole{label}{fig:tgv-run-time-errors}}
\end{figure}

Table \DUrole{ref}{table:weal-scaling} lists the wall time per 1 thousand iterations and the scaling efficiency.
As indicated previously, weak scaling was used in PINN, which follows most machine learning applications.\begin{table}
\setlength{\DUtablewidth}{\tablewidth}
\begin{longtable*}[c]{p{0.249\DUtablewidth}p{0.110\DUtablewidth}p{0.110\DUtablewidth}p{0.110\DUtablewidth}p{0.110\DUtablewidth}}
\toprule
 & \textbf{%
1 GPUs} & \textbf{%
2 GPUs} & \textbf{%
4 GPUs} & \textbf{%
8 GPUs} \\
\midrule
\endfirsthead

Time (sec/1k iters) & 
85.0 & 
87.7 & 
89.1 & 
90.1 \\

Efficiency (\%) & 
100 & 
97 & 
95 & 
94 \\
\bottomrule
\end{longtable*}
\caption{Weak scaling performance of the PINN solver using NVIDIA A100-80GB GPUs \DUrole{label}{table:weal-scaling}}\end{table}

\subsubsection{3.4. Discussion%
  \label{discussion}%
}

A notice should be made regarding the results: we do not claim that these results represent the most optimized configuration of the PINN method.
Neither do we claim the qualitative conclusions apply to all other hyperparameter configurations.
These results merely reflect the outcomes of our computational experiments with respect to the specific configuration abovementioned.
They should be deemed experimental data rather than a thorough analysis of the method's characteristics.

The Taylor-Green vortex serves as a good benchmark case because it reduces the number of required residual constraints: residuals $r_4$ and $r_5$ are excluded from $r$ in equation \DUrole{ref}{eq:total-residual}.
This means the optimizer can concentrate only on the residuals of initial conditions and the Navier-Stokes equations.

Using more GPUs (thus using more training points, i.e., spatio-temporal points) did not speed up the convergence, which may indicate that the per-iteration number of points on a single GPU is already big enough.
The number of training points mainly affects the mean gradients of the residual with respect to model parameters, which then will be used to update parameters by gradient-descent-based optimizers.
If the number of points is already big enough on a single GPU, then using more points or more GPUs is unlikely to change the mean gradients significantly, causing the convergence solely to rely on learning rates.

The accuracy of the PINN solver was acceptable but not satisfying, especially when considering how much time it took to achieve such accuracy.
The low accuracy to some degree was not surprising.
Recall the theory in section 2.
The PINN method only seeks the minimal residual on the total residual's hyperplane.
It does not try to find the zero root of the hyperplane and does not even care whether such a zero root exists.
Furthermore, by using a gradient-descent-based optimizer, the resulting minimum is likely just a local minimum.
It makes sense that it is hard for the residual to be close to zero, meaning it is hard to make errors small.

Regarding the performance result in figure \DUrole{ref}{fig:tgv-run-time-errors}, we would like to avoid interpreting the result as one solver being better than the other one.
The proper conclusion drawn from the figure should be as follows: when using the PINN solver as a CFD simulator for a specific flow condition, PetIBM outperforms the PINN solver.
As stated in section 1, the PINN method can solve flows under different flow parameters in one run—a capability that PetIBM does not have.
The performance result in figure \DUrole{ref}{fig:tgv-run-time-errors} only considers a limited application of the PINN solver.

One issue for this case study was how to fairly compare the PINN solver and PetIBM, especially when investigating the accuracy versus the workload/problem size or time-to-solution versus problem size.
Defining the problem size in PINN is not as straightforward as we thought.
Let us start with degrees of freedom—in PINN, it is called the number of model parameters, and in traditional CFD solvers, it is called the number of unknowns.
The PINN solver and traditional CFD solvers are all trying to determine the free parameters in models (that is, approximate solutions).
Hence, the number of degrees of freedom determines the problem sizes and workloads.
However, in PINN, problem sizes and workloads do not depend on degrees of freedom solely.
The number of training points also plays a critical role in workloads.
We were not sure if it made sense to define a problem size as the sum of the per-iteration number of training points and the number of model parameters.
For example, 100 model parameters plus 100 training points is not equivalent to 150 model parameters plus 50 training points in terms of workloads.
So without a proper definition of problem size and workload, it was not clear how to fairly compare PINN and traditional CFD methods.

Nevertheless, the gap between the performances of PINN and PetIBM is too large, and no one can argue that using other metrics would change the conclusion.
Not to mention that the PINN solver ran on A100 GPUs, while PetIBM ran on a single K40 GPU in our lab, a product from 2013.
This is also not a surprising conclusion because, as indicated in section 2, the use of automatic differentiation for temporal and spatial derivatives results in a huge computational graph.
In addition, the PINN solver uses gradient-descent based method, which is a first-order method and limits the performance.

Weak scaling is a natural choice of the PINN solver when it comes to distributed computing.
As we don't know a proper way to define workload, simply copying all model parameters to all processes and using the same number of training points on all processes works well.

\subsection{4. Case 2: 2D cylinder flows: harder than we thought%
  \label{case-2-2d-cylinder-flows-harder-than-we-thought}%
}

This case study shows what really made us frustrated: a 2D cylinder flow at Reynolds number $Re=200$.
We failed to even produce a solution that qualitatively captures the key physical phenomenon of this flow: vortex shedding.

\subsubsection{4.1. Problem description%
  \label{problem-description}%
}

The computational domain is $[-8, 25]\times[-8, 8]$, and a cylinder with a radius of $0.5$ sits at coordinate $(0, 0)$.
The velocity boundary conditions are $(u, v)=(1, 0)$ along $x=-8$, $y=-8$, and $y=8$.
On the cylinder surface is the no-slip condition, i.e., $(u, v)=(0, 0)$.
At the outlet ($x=25$), we enforced a pressure boundary condition $p=0$.
The initial condition is $(u, v)=(0, 0)$.
Note that this initial condition is different from most traditional CFD simulations.
Conventionally, CFD simulations use $(u, v)=(1, 0)$ for cylinder flows.
A uniform initial condition of $u=1$ does not satisfy the Navier-Stokes equations due to the no-slip boundary on the cylinder surface.
Conventional CFD solvers are usually able to correct the solution during time-marching by propagating boundary effects into the domain through numerical schemes' stencils.
In our experience, using $u=1$ or $u=0$ did not matter for PINN because both did not give reasonable results.
Nevertheless, the PINN solver's results shown in this section were obtained using a uniform $u=0$ for the initial condition.

The density, $\rho$, is one, and the kinematic viscosity is $\nu=0.005$.
These parameters correspond to Reynolds number $Re=200$.
Figure \DUrole{ref}{fig:cylinder-petibm-contour-t200} shows the velocity and vorticity snapshots at $t=200$.
As shown in the figure, this type of flow displays a phenomenon called vortex shedding.
Though vortex shedding makes the flow always unsteady, after a certain time, the flow reaches a periodic stage and the flow pattern repeats after a certain period.\begin{figure}[]\noindent\makebox[\columnwidth][c]{\includegraphics[width=\columnwidth]{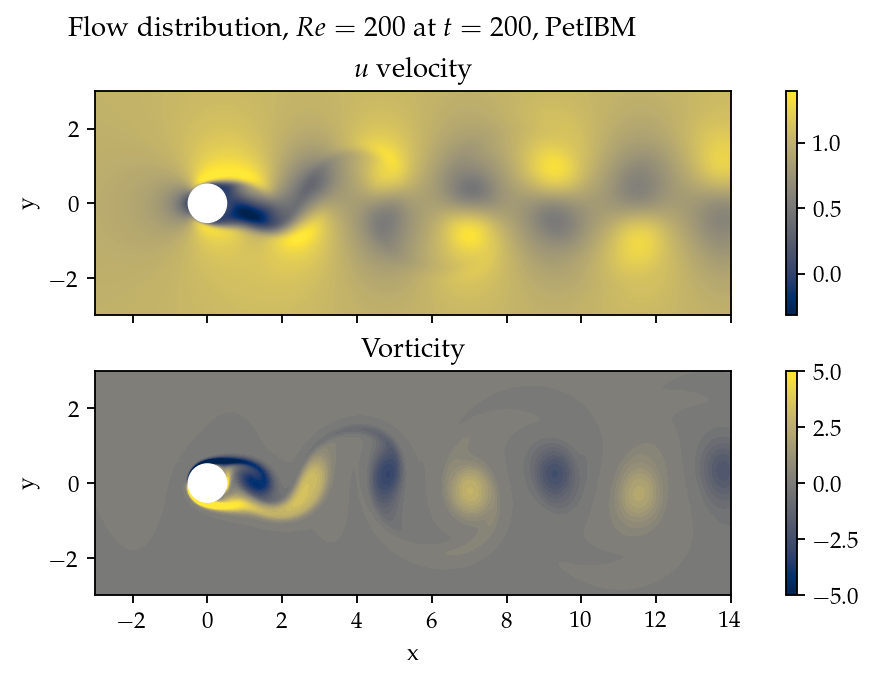}}
\caption{Demonstration of velocity and vorticity fields at $t=200$ from a PetIBM simulation. \DUrole{label}{fig:cylinder-petibm-contour-t200}}
\end{figure}

The Navier-Stokes equations can be deemed as a dynamical system.
Instability appears in the flow under some flow conditions and responds to small perturbations, causing the vortex shedding.
In nature, the vortex shedding comes from the uncertainty and perturbation existing everywhere.
In CFD simulations, the vortex shedding is caused by small numerical and rounding errors in calculations.
Interested readers should consult reference \DUrole{cite}{Williamson1996}.

\subsubsection{4.2. Solver and runtime configurations%
  \label{id1}%
}

For the PINN solver, we tested with two networks.
Both were fully-connected neural networks: one with 256 neurons per layer, while the other one with 512 neurons per layer.
All other network configurations were the same as those in section 3, except we allowed human intervention to manually adjust the learning rates during training.
Our intention for this case study was to successfully obtain physical solutions from the PINN solver, rather than conducting a performance and accuracy benchmark.
Therefore, we would adjust the learning rate to accelerate the convergence or to escape from local minimums.
This decision was in line with common machine learning practice.
We did not carry out hyperparameter optimization.
These parameters were chosen because they work in Modulus' examples and in the Taylor-Green vortex experiment.

The PINN solver pre-generated $40,960,000$ spatial-temporal points from a spatial domain in $[-8, 25]\times[-8, 8]$ and temporal domain $(0, 200]$ to evaluate residuals of the Navier-Stokes equations, and used $40,960$ points per iteration.
The number of pre-generated points for the initial condition was $2,048,000$, and the per-iteration number is $2,048$.
On each boundary, the numbers of pre-generated and per-iteration points are 8,192,000 and 8,192.
Both cases used 8 A100 GPUs, which scaled these numbers up with a factor of 8.
For example, during each iteration, a total of $327,680$ points were actually used to evaluate the Navier-Stokes equations' residuals.
Both cases ran up to 64 hours in wall time.

One PetIBM simulation was carried out as a baseline.
This simulation had a spatial resolution of $1485 \times 720$, and the time step size is 0.005.
Figure \DUrole{ref}{fig:cylinder-petibm-contour-t200} was rendered using this simulation.
The hardware used was 1 K40 GPU plus 6 cores of i7-5930K CPU.
It took about 1.7 hours to finish.

The quantity of interest is the drag coefficient.
We consider both the friction drag and pressure drag in the coefficient calculation as follows:\begin{equation}
\label{eq:drag-coefficient}
C_D=\frac{2}{\rho U_0^2 D}\int\limits_S\left(\rho\nu\frac{\partial \left(\vec{U}\cdot\vec{t}\right)}{\partial \vec{n}}n_y-pn_x\right)\mathrm{d}S
\end{equation}Here, $U_0=1$ is the inlet velocity. $\vec{n}=[n_x,n_y]^\mathsf{T}$ and $\vec{t}=[n_y, -n_x]^\mathsf{T}$ are the normal and tangent vectors, respectively.
$S$ represents the cylinder surface.
The theoretical lift coefficient ($C_L$) for this flow is zero due to the symmetrical geometry.

\subsubsection{4.3. Results%
  \label{id2}%
}

Note, as stated in section 3.4, we deem the results as experimental data under a specific experiment configuration.
Hence, we do not claim that the results and qualitative conclusions will apply to other hyperparameter configuration.

Figure \DUrole{ref}{fig:cylinder-pinn-training-convergence} shows the convergence history.
The bumps in the history correspond to our manual adjustment of the learning rates.
After 64 hours of training, the total loss had not converged to an obvious steady value.
However, we decided not to continue the training because, as later results will show, it is our judgment call that the results would not be correct even if the training converged.\begin{figure}[]\noindent\makebox[\columnwidth][c]{\includegraphics[width=\columnwidth]{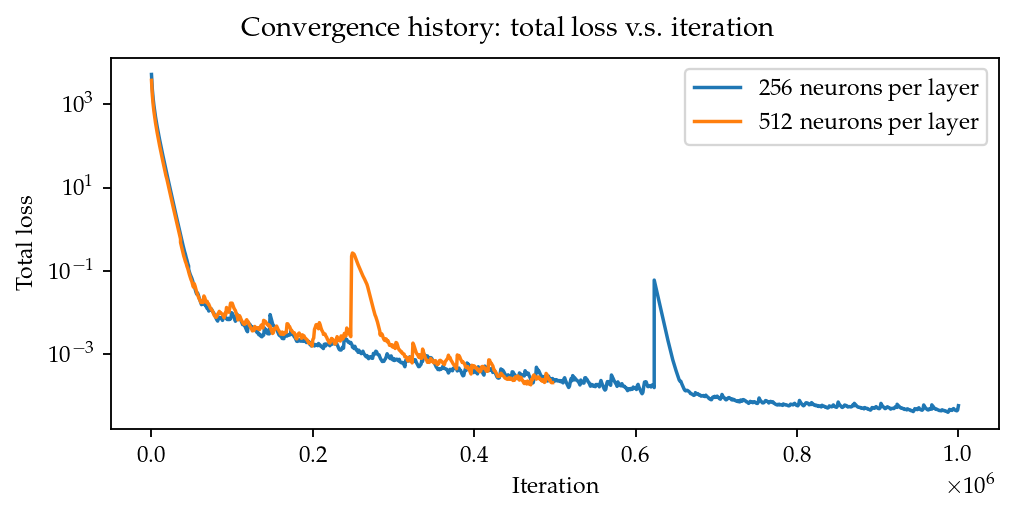}}
\caption{Training history of the 2D cylinder flow at $Re=200$. \DUrole{label}{fig:cylinder-pinn-training-convergence}}
\end{figure}

Figure \DUrole{ref}{fig:cylinder-pinn-contour-t200} provides a visualization of the predicted velocity and vorticity at $t=200$.
And in figure \DUrole{ref}{cylinder-cd-cl} are the drag and lift coefficients versus simulation time.
From both figures, we couldn't see any sign of vortex shedding with the PINN solver.\begin{figure}[]\noindent\makebox[\columnwidth][c]{\includegraphics[width=\columnwidth]{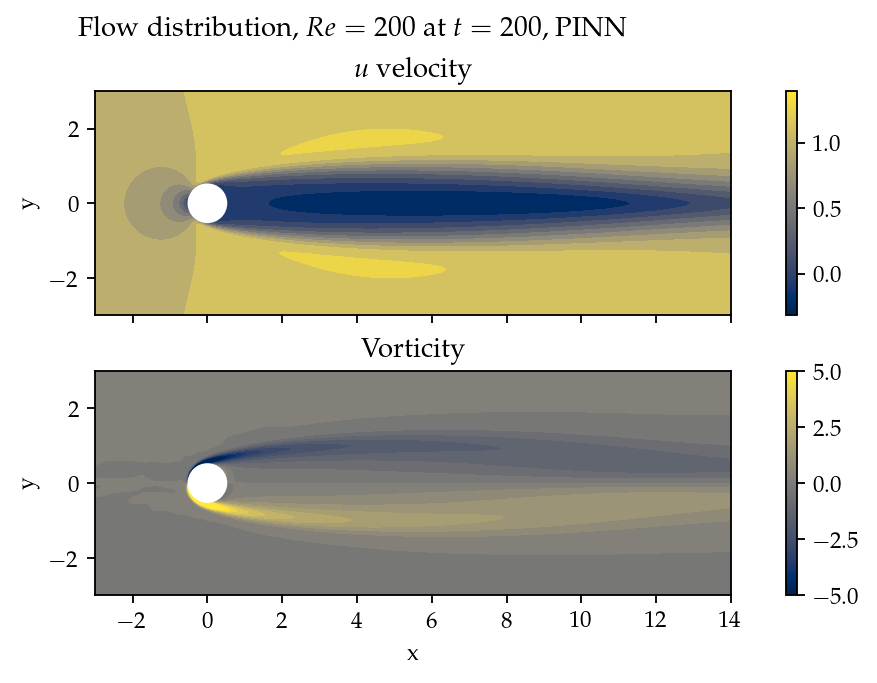}}
\caption{Velocity and vorticity at $t=200$ from PINN. \DUrole{label}{fig:cylinder-pinn-contour-t200}}
\end{figure}\begin{figure}[]\noindent\makebox[\columnwidth][c]{\includegraphics[width=\columnwidth]{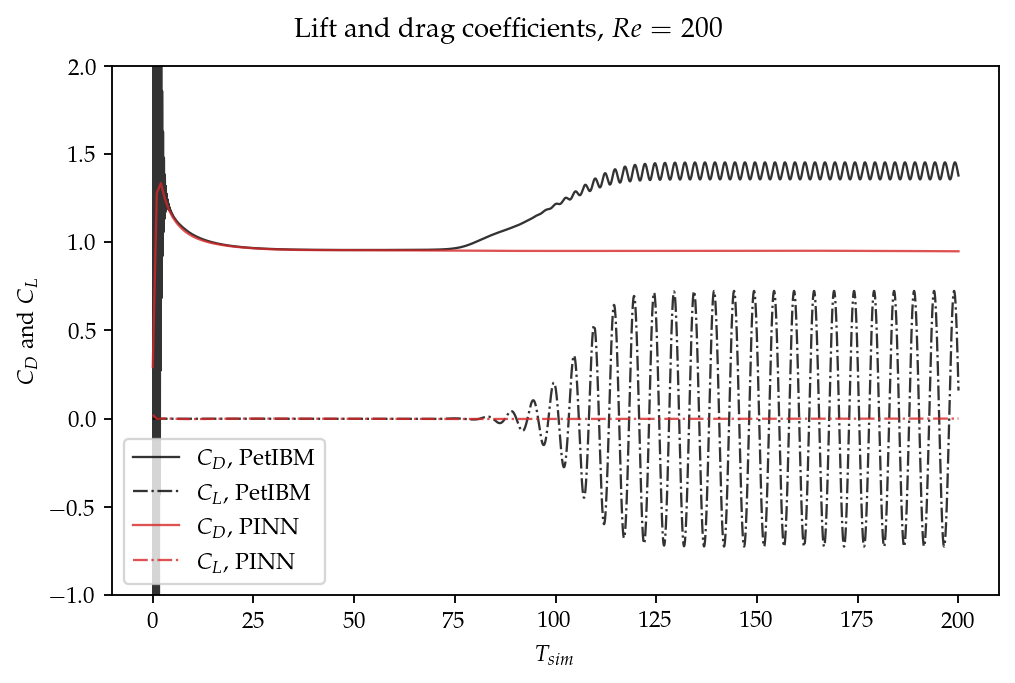}}
\caption{Drag and lift coefficients with respect to $t$ \DUrole{label}{cylinder-cd-cl}}
\end{figure}

We provide a comparison against the values reported by others in table \DUrole{ref}{table:drag-comparison}.
References \DUrole{cite}{gushchin_numerical_1974} and \DUrole{cite}{Fornberg1980} calculate the drag coefficients using steady flow simulations, which were popular decades ago because of their inexpensive computational costs.
The actual flow is not a steady flow, and these steady-flow coefficient values are lower than unsteady-flow predictions.
The drag coefficient from the PINN solver is closer to the steady-flow predictions.
\begin{table}
   \centering
   \begin{tabular}{cccccc}
      \toprule
      \multicolumn{2}{c}{} & \multicolumn{2}{c}{Unsteady simulations} & \multicolumn{2}{c}{Steady simulations} \\
      PetIBM & PINN & \cite{deng_hydrodynamic_2007} & \cite{Rajani2009} & \cite{gushchin_numerical_1974} & \cite{Fornberg1980} \\
      \midrule
      1.38 & 0.95 & 1.25 & 1.34 & 0.97 & 0.83 \\
      \bottomrule
   \end{tabular}
\caption{Comparison of drag coefficients, $C_D$}\label{table:drag-comparison}
\end{table}

\subsubsection{4.4. Discussion%
  \label{id3}%
}

While researchers may be interested in why the PINN solver behaves like a steady flow solver, in this section, we would like to focus more on the user experience and the usability of PINN in practice.
Our viewpoints may be subjective, and hence we leave them here in the discussion.

Allow us to start this discussion with a hypothetical situation.
If one asks why we chose such a spatial and temporal resolution for a conventional CFD simulation, we have mathematical or physical reasons to back our decision.
However, if the person asks why we chose 6 hidden layers and 256 neurons per layer, we will not be able to justify it.
\textquotedbl{}It worked in another case!\textquotedbl{} is probably the best answer we can offer.
The situation also indicates that we have systematic approaches to improve a conventional simulation but can only improve PINN's results through computer experiments.

Most traditional numerical methods have rigorous analytical derivations and analyses.
Each parameter used in a scheme has a meaning or a purpose in physical or numerical aspects.
The simplest example is the spatial resolution in the finite difference method, which controls the truncation errors in derivatives.
Or, the choice of the limiters in finite volume methods, used to inhibit the oscillation in solutions.
So when a conventional CFD solver produces unsatisfying or even non-physical results, practitioners usually have systematic approaches to identify the cause or improve the outcomes.
Moreover, when necessary, practitioners know how to balance the computational cost and the accuracy, which is a critical point for using computer-aided engineering.
Engineering always concerns the costs and outcomes.

On the other hand, the PINN method lacks well-defined procedures to control the outcome.
For example, we know the numbers of neurons and layers control the degrees of freedom in a model.
With more degrees of freedom, a neural network model can approximate a more complicated phenomenon.
However, when we feel that a neural network is not complicated enough to capture a physical phenomenon, what strategy should we use to adjust the neurons and layers?
Should we increase neurons or layers first?
By how much?

Moreover, when it comes to something non-numeric, it is even more challenging to know what to use and why to use it.
For instance, what activation function should we use and why?
Should we use the same activation everywhere?
Not to mention that we are not yet even considering a different network architecture here.

Ultimately, are we even sure that increasing the network's complexity is the right path?
Our assumption that the network is not complicated enough may just be wrong.

The following situation happened in this case study.
Before we realized the PINN solver behaved like a steady-flow solver, we attributed the cause to model complexity.
We faced the problem of how to increase the model complexity systematically.
Theoretically, we could follow the practice of the design of experiments (e.g., through grid search or Taguchi methods).
However, given the computational cost and the number of hyperparameters/options of PINN, a proper design of experiments is not affordable for us.
Furthermore, the design of experiments requires the outcome to change with changes in inputs.
In our case, the vortex shedding remains absent regardless of how we changed hyperparameters.

Let us move back to the flow problem to conclude this case study.
The model complexity may not be the culprit here.
Vortex shedding is the product of the dynamical systems of the Navier-Stokes equations and the perturbations from numerical calculations (which implicitly mimic the perturbations in nature).
Suppose the PINN solver's prediction was the steady-state solution to the flow.
We may need to introduce uncertainties and perturbations in the neural network or the training data, such as a perturbed initial condition described in \DUrole{cite}{laroussi_vortex_2015}.
As for why PINN predicts the steady-state solution, we cannot answer it currently.

\subsection{5. Further discussion and conclusion%
  \label{further-discussion-and-conclusion}%
}

Because of the widely available deep learning libraries, such as PyTorch, and the ease of Python, implementing a PINN solver is relatively more straightforward nowadays.
This may be one reason why the PINN method suddenly became so popular in recent years.
This paper does not intend to discourage people from trying the PINN method.
Instead, we share our failures and frustration using PINN so that interested readers may know what immediate challenges should be resolved for PINN.

Our paper is limited to using the PINN solver as a replacement for traditional CFD solvers.
However, as the first section indicates, PINN can do more than solving one specific flow under specific flow parameters.
Moreover, PINN can also work with traditional CFD solvers.
The literature shows researchers have shifted their attention to hybrid-mode applications.
For example, in \DUrole{cite}{jiang_meshfreeflownet_2020}, the authors combined the concept of PINN and a traditional CFD solver to train a model that takes in low-resolution CFD simulation results and outputs high-resolution flow fields.

For people with a strong background in numerical methods or CFD, we would suggest trying to think out of the box.
During our work, we realized our mindset and ideas were limited by what we were used to in CFD.
An example is the initial conditions.
We were used to only having one set of initial conditions when the temporal derivative in differential equations is only first-order.
However, in PINN, nothing limits us from using more than one initial condition.
We can generate results at $t=0,1,\dots,t_n$ using a traditional CFD solver and add the residuals corresponding to these time snapshots to the total residual, so the PINN method may perform better in predicting $t>t_n$.
In other words, the PINN solver becomes the traditional CFD solvers' replacement only for $t>t_n$ (\DUrole{cite}{noauthor_modulus_nodate}).

As discussed in \DUrole{cite}{thuerey_physics-based_2022}, solving partial differential equations with deep learning is still a work-in-progress.
It may not work in many situations.
Nevertheless, it does not mean we should stay away from PINN and discard this idea.
Stepping away from a new thing gives zero chance for it to evolve, and we will never know if PINN can be improved to a mature state that works well.
Of course, overly promoting its bright side with only success stories does not help, either.
Rather, we should honestly face all troubles, difficulties, and challenges.
Knowing the problem is the first step to solving it.

\subsection{Acknowledgements%
  \label{acknowledgements}%
}

We appreciate the support by NVIDIA, through sponsoring the access to its high-performance computing cluster.
\bibliographystyle{alphaurl}
\bibliography{reference}

\end{document}

%% file: page_numbers.tex
\setcounter{page}{1}

%% file: paper.bbl
\newcommand{\etalchar}[1]{$^{#1}$}
\begin{thebibliography}{CMKAB18}

\bibitem[Chu22]{pi_yueh_chuang_2022_6592457}
Pi-Yueh Chuang.
\newblock barbagroup/scipy-2022-repro-pack: 20220530, May 2022.
\newblock \href {https://doi.org/10.5281/zenodo.6592457}
  {\path{doi:10.5281/zenodo.6592457}}.

\bibitem[CMKAB18]{chuang_petibm_2018}
Pi-Yueh Chuang, Olivier Mesnard, Anush Krishnan, and Lorena A.~Barba.
\newblock {PetIBM}: toolbox and applications of the immersed-boundary method on
  distributed-memory architectures.
\newblock {\em Journal of Open Source Software}, 3(25):558, May 2018.
\newblock URL: \url{http://joss.theoj.org/papers/10.21105/joss.00558}, \href
  {https://doi.org/10.21105/joss.00558} {\path{doi:10.21105/joss.00558}}.

\bibitem[CMW{\etalchar{+}}]{cai_physics-informed_2021}
Shengze Cai, Zhiping Mao, Zhicheng Wang, Minglang Yin, and George~Em
  Karniadakis.
\newblock Physics-informed neural networks ({PINNs}) for fluid mechanics: a
  review.
\newblock 37(12):1727--1738.
\newblock URL: \url{https://link.springer.com/10.1007/s10409-021-01148-1},
  \href {https://doi.org/10.1007/s10409-021-01148-1}
  {\path{doi:10.1007/s10409-021-01148-1}}.

\bibitem[DPT]{dissanayake_neural-network-based_1994}
M.~W. M.~G. Dissanayake and N.~Phan-Thien.
\newblock Neural-network-based approximations for solving partial differential
  equations.
\newblock 10(3):195--201.
\newblock URL:
  \url{https://onlinelibrary.wiley.com/doi/10.1002/cnm.1640100303}, \href
  {https://doi.org/10.1002/cnm.1640100303} {\path{doi:10.1002/cnm.1640100303}}.

\bibitem[DSY07]{deng_hydrodynamic_2007}
Jian Deng, Xue-Ming Shao, and Zhao-Sheng Yu.
\newblock Hydrodynamic studies on two traveling wavy foils in tandem
  arrangement.
\newblock {\em Physics of Fluids}, 19(11):113104, November 2007.
\newblock URL: \url{http://aip.scitation.org/doi/10.1063/1.2814259}, \href
  {https://doi.org/10.1063/1.2814259} {\path{doi:10.1063/1.2814259}}.

\bibitem[DZ]{du_evolutional_2021}
Yifan Du and Tamer~A. Zaki.
\newblock Evolutional deep neural network.
\newblock 104(4):045303.
\newblock URL: \url{https://link.aps.org/doi/10.1103/PhysRevE.104.045303},
  \href {https://doi.org/10.1103/PhysRevE.104.045303}
  {\path{doi:10.1103/PhysRevE.104.045303}}.

\bibitem[For80]{Fornberg1980}
Bengt Fornberg.
\newblock A numerical study of steady viscous flow past a circular cylinder.
\newblock {\em Journal of Fluid Mechanics}, 98(04):819, June 1980.
\newblock URL:
  \url{http://www.journals.cambridge.org/abstract_S0022112080000419}, \href
  {https://doi.org/10.1017/S0022112080000419}
  {\path{doi:10.1017/S0022112080000419}}.

\bibitem[FS]{Faller1997}
William~E. Faller and Scott~J. Schreck.
\newblock Unsteady fluid mechanics applications of neural networks.
\newblock 34(1):48--55.
\newblock URL: \url{http://arc.aiaa.org/doi/10.2514/2.2134}, \href
  {https://doi.org/10.2514/2.2134} {\path{doi:10.2514/2.2134}}.

\bibitem[GS74]{gushchin_numerical_1974}
V.A. Gushchin and V.V. Shchennikov.
\newblock A numerical method of solving the navier-stokes equations.
\newblock {\em USSR Computational Mathematics and Mathematical Physics},
  14(2):242--250, January 1974.
\newblock URL:
  \url{https://linkinghub.elsevier.com/retrieve/pii/0041555374900615}, \href
  {https://doi.org/10.1016/0041-5553(74)90061-5}
  {\path{doi:10.1016/0041-5553(74)90061-5}}.

\bibitem[Hao]{hao_ai_2020}
Karen Hao.
\newblock {AI} has cracked a key mathematical puzzle for understanding our
  world.
\newblock URL:
  \url{https://www.technologyreview.com/2020/10/30/1011435/ai-fourier-neural-network-cracks-navier-stokes-and-partial-differential-equations/}.

\bibitem[HG]{hendrycks_gaussian_2016}
Dan Hendrycks and Kevin Gimpel.
\newblock Gaussian error linear units ({GELUs}).
\newblock Publisher: {arXiv} Version Number: 4.
\newblock URL: \url{https://arxiv.org/abs/1606.08415}, \href
  {https://doi.org/10.48550/ARXIV.1606.08415}
  {\path{doi:10.48550/ARXIV.1606.08415}}.

\bibitem[Hor]{hornik_approximation_1991}
Kurt Hornik.
\newblock Approximation capabilities of multilayer feedforward networks.
\newblock 4(2):251--257.
\newblock URL:
  \url{https://linkinghub.elsevier.com/retrieve/pii/089360809190009T}, \href
  {https://doi.org/10.1016/0893-6080(91)90009-T}
  {\path{doi:10.1016/0893-6080(91)90009-T}}.

\bibitem[JEA{\etalchar{+}}20]{jiang_meshfreeflownet_2020}
Chiyu~“Max” Jiang, Soheil Esmaeilzadeh, Kamyar Azizzadenesheli, Karthik
  Kashinath, Mustafa Mustafa, Hamdi~A. Tchelepi, Philip Marcus, Mr~Prabhat, and
  Anima Anandkumar.
\newblock Meshfreeflownet: A physics-constrained deep continuous space-time
  super-resolution framework.
\newblock In {\em SC20: International Conference for High Performance
  Computing, Networking, Storage and Analysis}, pages 1--15, 2020.
\newblock \href {https://doi.org/10.1109/SC41405.2020.00013}
  {\path{doi:10.1109/SC41405.2020.00013}}.

\bibitem[KDYI]{karali_novel_2021}
Hasan Karali, Umut~M. Demirezen, Mahmut~A. Yukselen, and Gokhan Inalhan.
\newblock A novel physics informed deep learning method for simulation-based
  modelling.
\newblock In {\em {AIAA} Scitech 2021 Forum}. American Institute of Aeronautics
  and Astronautics.
\newblock URL: \url{https://arc.aiaa.org/doi/10.2514/6.2021-0177}, \href
  {https://doi.org/10.2514/6.2021-0177} {\path{doi:10.2514/6.2021-0177}}.

\bibitem[LD15]{laroussi_vortex_2015}
Mouna Laroussi and Mohamed Djebbi.
\newblock Vortex {Shedding} for {Flow} {Past} {Circular} {Cylinder}: {Effects}
  of {Initial} {Conditions}.
\newblock {\em Universal Journal of Fluid Mechanics}, 3:19--32, 2015.

\bibitem[LLF]{lagaris_artificial_1998}
I.~E. Lagaris, A.~Likas, and D.~I. Fotiadis.
\newblock Artificial neural networks for solving ordinary and partial
  differential equations.
\newblock 9(5):987--1000.
\newblock URL: \url{http://ieeexplore.ieee.org/document/712178/}, \href
  {http://arxiv.org/abs/physics/9705023} {\path{arXiv:physics/9705023}}, \href
  {https://doi.org/10.1109/72.712178} {\path{doi:10.1109/72.712178}}.

\bibitem[LLQH]{Li2003}
Jianyu Li, Siwei Luo, Yingjian Qi, and Yaping Huang.
\newblock Numerical solution of elliptic partial differential equation using
  radial basis function neural networks.
\newblock 16(5):729--734.
\newblock URL:
  \url{https://linkinghub.elsevier.com/retrieve/pii/S0893608003000832}, \href
  {https://doi.org/10.1016/S0893-6080(03)00083-2}
  {\path{doi:10.1016/S0893-6080(03)00083-2}}.

\bibitem[LMMK]{lu_deepxde:_2021}
Lu~Lu, Xuhui Meng, Zhiping Mao, and George~Em Karniadakis.
\newblock {DeepXDE}: A deep learning library for solving differential
  equations.
\newblock 63(1):208--228.
\newblock URL: \url{https://epubs.siam.org/doi/10.1137/19M1274067}, \href
  {https://doi.org/10.1137/19M1274067} {\path{doi:10.1137/19M1274067}}.

\bibitem[LS]{Linse1993}
Dennis~J. Linse and Robert~F. Stengel.
\newblock Identification of aerodynamic coefficients using computational neural
  networks.
\newblock 16(6):1018--1025.
\newblock Publisher: Springer {US}, Place: Boston, {MA}.
\newblock URL: \url{http://link.springer.com/10.1007/0-306-48610-5_9}, \href
  {https://doi.org/10.2514/3.21122} {\path{doi:10.2514/3.21122}}.

\bibitem[noa]{noauthor_modulus_nodate}
Modulus.
\newblock URL: \url{https://docs.nvidia.com/deeplearning/modulus/index.html}.

\bibitem[RKM09]{Rajani2009}
B.N. Rajani, A.~Kandasamy, and Sekhar Majumdar.
\newblock Numerical simulation of laminar flow past a circular cylinder.
\newblock {\em Applied Mathematical Modelling}, 33(3):1228--1247, March 2009.
\newblock arXiv: DOI: 10.1002/fld.1 Publisher: Elsevier Inc. ISBN: 02712091
  10970363.
\newblock URL: \url{http://dx.doi.org/10.1016/j.apm.2008.01.017}, \href
  {https://doi.org/10.1016/j.apm.2008.01.017}
  {\path{doi:10.1016/j.apm.2008.01.017}}.

\bibitem[RPK]{raissi_physics-informed_2019}
M.~Raissi, P.~Perdikaris, and G.E. Karniadakis.
\newblock Physics-informed neural networks: A deep learning framework for
  solving forward and inverse problems involving nonlinear partial differential
  equations.
\newblock 378:686--707.
\newblock URL:
  \url{https://linkinghub.elsevier.com/retrieve/pii/S0021999118307125}, \href
  {https://doi.org/10.1016/j.jcp.2018.10.045}
  {\path{doi:10.1016/j.jcp.2018.10.045}}.

\bibitem[SS]{sirignano_dgm:_2018}
Justin Sirignano and Konstantinos Spiliopoulos.
\newblock {DGM}: A deep learning algorithm for solving partial differential
  equations.
\newblock 375:1339--1364.
\newblock URL:
  \url{https://linkinghub.elsevier.com/retrieve/pii/S0021999118305527}, \href
  {https://doi.org/10.1016/j.jcp.2018.08.029}
  {\path{doi:10.1016/j.jcp.2018.08.029}}.

\bibitem[THM{\etalchar{+}}]{thuerey_physics-based_2022}
Nils Thuerey, Philipp Holl, Maximilian Mueller, Patrick Schnell, Felix Trost,
  and Kiwon Um.
\newblock Physics-based deep learning.
\newblock Number: {arXiv}:2109.05237.
\newblock URL: \url{http://arxiv.org/abs/2109.05237}, \href
  {http://arxiv.org/abs/2109.05237 [physics]} {\path{arXiv:2109.05237
  [physics]}}.

\bibitem[Tre]{trefethen_spectral_2000}
Lloyd~N. Trefethen.
\newblock {\em Spectral Methods in {MATLAB}}.
\newblock Software, environments, tools. Society for Industrial and Applied
  Mathematics.
\newblock URL: \url{http://epubs.siam.org/doi/book/10.1137/1.9780898719598},
  \href {https://doi.org/10.1137/1.9780898719598}
  {\path{doi:10.1137/1.9780898719598}}.

\bibitem[Wil]{Williamson1996}
C.~H.~K. Williamson.
\newblock Vortex dynamics in the cylinder wake.
\newblock 28(1):477--539.
\newblock URL:
  \url{http://www.annualreviews.org/doi/10.1146/annurev.fl.28.010196.002401},
  \href {https://doi.org/10.1146/annurev.fl.28.010196.002401}
  {\path{doi:10.1146/annurev.fl.28.010196.002401}}.

\bibitem[WTP]{wang_understanding_2021}
Sifan Wang, Yujun Teng, and Paris Perdikaris.
\newblock Understanding and mitigating gradient flow pathologies in
  physics-informed neural networks.
\newblock 43(5):A3055--A3081.
\newblock URL: \url{https://epubs.siam.org/doi/10.1137/20M1318043}, \href
  {https://doi.org/10.1137/20M1318043} {\path{doi:10.1137/20M1318043}}.

\bibitem[WYP]{wang_when_2022}
Sifan Wang, Xinling Yu, and Paris Perdikaris.
\newblock When and why {PINNs} fail to train: A neural tangent kernel
  perspective.
\newblock 449:110768.
\newblock URL:
  \url{https://linkinghub.elsevier.com/retrieve/pii/S002199912100663X}, \href
  {https://doi.org/10.1016/j.jcp.2021.110768}
  {\path{doi:10.1016/j.jcp.2021.110768}}.

\end{thebibliography}
